\documentclass[aps,nofootinbib,preprint,groupedaddress]{revtex4}
\usepackage{graphicx}
\newcommand{\be}{\begin{equation}}
\newcommand{\ee}{\end{equation}}
\newcommand{\bear}{\begin{eqnarray}}
\newcommand{\eear}{\end{eqnarray}}
\begin{document}

\title{The nature of power corrections in large $\beta_0$ approximation}

\author{Taekoon Lee}
\email{tlee@phya.snu.ac.kr}

\affiliation{Department of Physics, Seoul National University, Seoul 151-742,
Korea}


\begin{abstract}
We investigate the nature of power corrections and  infrared
renormalon singularities in large $\beta_0$ approximation. We argue that
the power correction associated with a renormalon pole singularity
should appear at $O(1)$,  in contrast to the renormalon 
ambiguity appearing at $O(1/\beta_0)$, 
and give an explanation why the leading order 
renormalon singularities are generically
{\it poles}.

\end{abstract}



\maketitle

The perturbative expansion in weak coupling constant is in general
an asymptotic expansion, with the perturbative coefficients growing
factorially at large orders. When the large order behavior is sign
alternating the series may be Borel resummed. On the other hand,
a large order behavior of same sign generally implies presence of
nonperturbative effects, and in such a case Borel resummation of the
series becomes ambiguous. 

According to the heuristic but general argument of Borel resummation
of a series with a same sign large order behavior \cite{justin}, the series 
is first Borel resummed at an unphysical negative coupling, at which
it becomes sign alternating, and then analytically continued 
back to the positive physical coupling, to yield a resummed amplitude.
It turns out that the resummed amplitude obtained in this way
has a cut along the positive real axis in the complex coupling plane,
and consequently has an ambiguous imaginary part at physical coupling.
This unphysical imaginary part is then supposed to be canceled by adding
a nonperturbative effect, and  the total amplitude is given as the sum of
the respective real parts of the Borel resummed amplitude and  
the nonperturbative effect.

In this brief note we study the nature of the power corrections 
in large $\beta_0$ approximation \cite{largebeta}, in which the number
of quark flavors $N_f$ (or equivalently $\beta_0 (\sim N_f)$, 
the first coefficient of QCD beta function), is pushed to (negative) infinity 
while $ a_s\equiv N_f\alpha_s$ is held fixed, where $\alpha_s$ denotes 
the strong coupling. 
A characteristic feature of the
Borel transform of an amplitude  in this scheme is that all known
renormalon singularities at leading order are poles 
(in general renormalon singularities are
branch cuts), and the Borel transform is subleading
in $1/\beta_0$ expansion, vanishing at $\beta_0\to\infty$.
To the author's knowledge there is no general explanation for the former,
except within the special context of a particular calculation, but
the latter comes from the very definition of the approximation. Since
the leading term in $a_s$ expansion comes from a gluon exchange with no
fermion bubble inserted, the large $\beta_0$ perturbative expansion of
a generic amplitude  $A(\alpha_s)$ takes the form
\be
A(\alpha_s)= \frac{1}{\beta_0} \sum_{n=0}^\infty a_n a_s^{n+1}\,,
\ee
where $a_n$ are $\beta_0$ independent,
which gives the Borel transform
\be
\tilde A(b) =\frac{1}{\beta_0} \sum_{n=0}^\infty \frac{a_n}{n!} b^n\equiv
\frac{1}{\beta_0} f(b)\,.
\ee
Consequently, the renormalon ambiguity of $A(\alpha_s)$ due to a
pole in $f(b)$ is $O(1/\beta_0)$, and vanishes at $\beta_0\to\infty$. 
In the following we shall see that these features are simply two sides of the
same coin.

We begin by observing an universality present in the 
nonperturbative effect.
Since the Borel resummation of a same sign series is based on a general, 
model independent argument, the term in the nonperturbative effect 
that gives rise to 
the cut singularity along the positive real axis in the coupling plane 
should be universal for a given large order behavior, or equivalently
for a given renormalon singularity in Borel plane, and is independent of the
specifics of the underlying theory. Thus, the power correction 
of a pole singularity
in large $\beta_0$ approximation must share a common functional form with
the nonperturbative effect associated with the same type pole
singularity in other models, for instance, such solvable models as 
the quantum mechanical double well potential or the  two-dimensional
$O(N)$ nonlinear sigma model. In these solvable models the ambiguous term in
the nonperturbative effect associated with a pole  in Borel plane,
\be
\frac{1}{(1-b/b_0)^{1+m}}, \hspace{.25in}\mbox{$m$ a non-negative integer}\,,
\label{e3}
\ee
assumes the form \cite{lee-nonp}
\be
\frac{b_0^{1+m}}{m!} e^{-b_0/\alpha} \alpha^{-m} \ln (-\alpha)\,,
\label{e4}
\ee
where $\alpha$ denotes the coupling constant in consideration, 
and we expect the power correction of a renormalon pole
in large $\beta_0$ approximation
would also contain a term of the same functional form as (\ref{e4}). Note that
the cut along the positive axis is provided by
the log function. The correspondence between Eq. (\ref{e3}) and  (\ref{e4})
can be easily confirmed using
the relation 
\be
A_{\rm PT} (\alpha\pm i\epsilon) = \int_{0\pm i\epsilon}^{\infty \pm
i\epsilon} e^{-\frac{b}{\alpha}} \tilde A(b)\, db\,,
\ee
where $A_{\rm PT}$ denotes the Borel resummed perturbative contribution, and
the cancellation of the respective imaginary parts 
in $A_{\rm PT}$ and the nonperturbative effect.

To see how a nonperturbative term of this type arises in
large $\beta_0$ approximation we consider the power correction in
full QCD. From the renormalization group equation for the power correction
of dimension $n$ operator the power correction, denoted by 
$A_{\rm NP}$, may be written
as
\be
A_{\rm NP}(\alpha_s)
\propto \alpha_s^{-\nu} e^{-\frac{n}{2\beta_0\alpha_s}} 
[1+O(\alpha_s)]\,,
\ee
where
\be
\nu= n\frac{ \beta_1}{2\beta_0^2} +\frac{\gamma_0}{\beta_0}
\ee
and
\bear
\beta_0=\frac{1}{4\pi}(11-\frac{2}{3}N_f)\,, \hspace{.25in}
\beta_1=\frac{1}{(4\pi)^2}(102-\frac{38}{3}N_f)\,.
\eear
Here $\beta_1$ and $\gamma_0$ are
the second coefficient of the beta function and the 
one loop anomalous dimension of the dimension $n$ operator, respectively.
Since a cut can arise only from the pre-exponential factor the ambiguous
term in the power
correction, denoted by $A_{\rm NP}^{{\rm amb.}}$, may be written as
\be
A_{\rm NP}^{{\rm amb.}}= C  (-a_s)^{-\nu} e^{-\frac{n}{2a_s}} 
[1+\frac{1}{\beta_0}O(a_s)]\,,
\label{e8}
\ee
where $C$ is a real constant that may depend on $\beta_0$.
We now make an observation that plays a crucial role in the following, namely,
that for this power correction  
to be well-defined  at $\beta_0\to\infty$,
$C$ must be bounded in the same limit. 
Thus $C$ may be written as
\be
C= c_0 + O(1/\beta_0)\,,
\ee
where $c_0$ is a $\beta_0$ independent constant. 

Then the ambiguous imaginary part due to the branch cut 
in the power correction 
is given by
\be
{\rm Im} A_{\rm NP}^{{\rm amb.}} (a_s\pm i\epsilon)= \pm c_0 \sin(\nu\pi)
e^{-\frac{n}{2a_s}} a_s^{-\nu}[1 +O(1/\beta_0)]\,.
\ee
For this to be
$O(1/\beta_0)$,  we have
\be
c_0 \sin(\nu\pi) \sim O(1/\beta_0)\,,
\ee
which allows one to write $\nu$ as
\be
\nu =\kappa +\frac{\chi}{\beta_0} + O(1/\beta_0^2)\,,
\label{e12}
\ee
where $\kappa$ is an integer and $\chi$ is a real constant,
and
\be
c_0\sin(\nu\pi)=\frac{(-1)^\kappa c_0\chi\pi}{\beta_0}[1+O(1/\beta_0)]\,.
\ee
With Eq. (\ref{e12}),
the expansion of the power correction in $1/\beta_0$ gives
\be
A_{\rm NP}^{{\rm amb.}}= (-1)^\kappa c_0 e^{-\frac{n}{2 a_s}} a_s^{-\kappa}
[ 1- \frac{\chi}{\beta_0}\ln(-a_s) +O(1/\beta_0^2)]\,,
\ee
which contains the logarithmic cut of precisely the type (\ref{e4})
that we were looking for.
The corresponding renormalon singularity is a pole,
\be
\tilde A(b) = \frac{c_0\chi\kappa! \left(\frac{-2}{n}\right)^{1+\kappa}
}{\beta_0} \frac{1}{(1-2 b/n)^{1+\kappa}} [1+
O(1-2b/n)]\,.
\label{e15}
\ee
This shows that the  $1/\beta_0$ suppression of
the renormalon ambiguity and  the pole nature of
the renormalon singularity share the same origin and 
are just two sides of the same coin.

Now taking the real and imaginary parts of the power correction we have,
to $O(1/\beta_0)$,
\bear
{\rm Re}\, A_{\rm NP}^{{\rm amb.}}(a_s\pm i\epsilon)&=&
(-1)^\kappa c_0 e^{-\frac{n}{2a_s}} a_s^{-\kappa}
[ 1-\frac{\chi}{\beta_0}\ln(a_s) ]\,, \nonumber\\
{\rm Im}\, A_{\rm NP}^{{\rm amb.}}(a_s\pm i\epsilon)&=&
\pm\frac{(-1)^\kappa c_0\pi\chi}{\beta_0} e^{-\frac{n}{2a_s}}
a_s^{-\kappa} \,.
\eear
Note that the imaginary part is subleading to the real part, as we claimed,
and the functional form for the real part to $O(1/\beta_0)$  is different from
that of the imaginary part. 
This shows that the common practice of taking the renormalon ambiguity
(imaginary part)  as an 
estimate of the magnitude of the power correction \footnote {For example, Ref.
\cite{Beneke:1998ui} takes $\frac{1}{\pi}$ of the imaginary part as
the nonperturbative effect in the case of Adler function. See also
\cite{sumino}.} is not only 
incorrect functionally
but also misses the leading  contribution which, having no cut, is 
invisible to the perturbation theory. This missing of the leading power 
correction by  perturbation
theory is not unique to this example; It occurs in other models with a pole
singularity, for instance, in $1/N$ expansion of the
two-dimensional $O(N)$ nonlinear sigma model \cite{sigmamodel}, in which
the sigma self energy has
a dynamically generated mass at $O[(1/N)^0]$, which,  being free of cut
singularity in coupling plane, does not cause any divergent large order 
behavior. In this model the perturbative weak coupling expansion 
starts at $O(1/N)$,
and the renormalon ambiguity, which is necessarily of $O(1/N)$, cannot
trace the leading order dynamical mass. Essentially the
same thing happens with the instanton-antiinstanton caused nonperturbative 
effect in the double well model, where the cut-free nonperturbative term 
cannot be traced by perturbation theory \cite{lee-nonp}.

So far, we  assumed  implicitly  that $c_0$  is a non-zero constant.
This assumption is natural since there is no reason that $c_0$ should
vanish. However, if $c_0$ is vanishing and $C$ is $O(1/\beta_0)$, 
then it is easy to see that $\nu$ in Eq. (\ref{e8})
must be a non-integer, which implies that the renormalon singularity should be
a cut rather than a pole, and that the real part of the nonperturbative 
effect occur at the same order as the imaginary part. In
large $\beta_0$ approximation there is no known
renormalon singularity of this type, and we conclude that this possibility is
not realized and $c_0$ be a nonvanishing constant.

Finally, we apply the discussion so far to an explicit example,
the power correction by the dimension 
four gluon condensate
in the Adler function $D[\alpha_s]$.
In this case the anomalous dimension vanishes, 
and we have
\be
\nu = 2 \frac{\beta_1}{\beta_0^2}
= \frac{2}{\beta_0} \frac{19}{4\pi} +O(1/\beta_0^2)\,, 
\ee
which is indeed in the form (\ref{e12}) with
\be
\kappa=0\,, \hspace{.25in} \chi= \frac{19}{2\pi}\,.
\label{e16}\ee
The exact Borel transform in the $\overline{\rm MS}$ scheme 
of the Adler function in large $\beta_0$
approximation is given as \cite{Beneke:1998ui,Broadhurst} 
\be
\tilde D(b) = \frac{1}{\pi\beta_0} 
\frac{e^{\frac{10}{3}}}{(1-b/2)}[1+O(1-b/2)]\,.
\ee
Comparing this with  Eq. (\ref{e15}), with $\kappa, \chi$ given in Eq.
(\ref{e16}),
we find
\be
c_0=-\frac{4}{19} e^{\frac{10}{3}}\,,
\ee
and the real part of the power correction
\be
{\rm Re} \,D_{\rm NP}^{{\rm amb.}}(\alpha_s) = c_0 e^{-\frac{2}{a_s}} 
[ 1-\frac{19}{2\pi\beta_0}\ln(a_s)]\,,
\ee
which is nonvanishing in $\beta_0\to\infty$.
This example shows that the physically relevant real part is indeed $O(1)$
compared to the $O(1/\beta_0)$ imaginary part, and  the cancellation
of the ambiguities in the Borel resummed amplitude and nonperturbative
effect allows one to recover part of the nonperturbative effect, even
those not visible perturbatively, from perturbation theory.

To conclude, based on a general argument on Borel resummation
of a same sign large order behavior, we have shown that in general
the renormalon singularities in large $\beta_0$ approximation should be  poles,
and the power corrections be of $O(1)$ whereas the renormalon ambiguities are
$O(1/\beta_0)$.

\begin{acknowledgements}
The author is thankful to the support from National Center for Theoretical
Sciences, Hsinchu, Taiwan.

\end{acknowledgements}


\end{document}